# Interaction of Global-Scale Atmospheric Vortices: Modeling Based on Hamiltonian System for Antipodal Vortices on Rotating Sphere

Igor I. Mokhov[*a], Sergey G. Chefranov[a], and Alexander G. Chefranov[b]

[a] *A.M. Obukhov Institute of Atmospheric Physics RAS, Moscow, Russia*
[b] *Eastern Mediterranean University, Famagusta, North Cyprus*

**Abstract**

It is shown for the first time that only an antipodal vortex pair (APV) is the elementary singular vortex object on the sphere compatible with the hydrodynamic equations. The exact weak solution of the absolute vorticity equation on the rotating sphere is obtained in the form of Hamiltonian dynamic system for $N$ interacting APVs. This is the first model describing interaction of Barrett vortices corresponding to atmospheric centers of action (ACA). In particular, new steady-state conditions for $N=2$ are obtained. These analytical conditions are used for the analysis of coupled cyclone-anticyclone ACAs over oceans in the Northern Hemisphere.

## 1. Introduction

Investigation of the fluid dynamics on the rotating sphere has fundamental and applied significance for understanding of dominant physical processes in the atmosphere and ocean. Development of such investigations on the base of wave and vortex representations and models is conducted mostly independently. All of them as a rule are based on solving of one and the same equation of absolute vorticity conservation in a thin layer of ideal incompressible fluid on the surface of a rotating sphere.

For corresponding investigations, two directions can be selected. The first one is related with the consideration of the point vortices on the sphere basics of which were founded by Gromeka [1], Zermelo [2] and others. The second direction was initiated by Rossby and is related with the use of Rossby waves [3]. In this paper we present results which are linked, though based on the dynamics modeling of point vortices on the sphere, with results of Barrett [4] developing studies initiated by Rossby. For better structuring of the obtained results we give in Section 2 some conclusions of studies on the base of Rossby waves and Barrett vortices. Their modified representation adequate to obtained in Section 3 and further new results is also given. The latter can be used for mode ling of quasi-stationary modes and their stability, in particular, for atmospheric centers of action (ACA) using a new approach for description of point vortices on the rotating sphere. The present paper is a more complete version of previous authors' publications [5,6].

## 2. Rossby waves and Barrett vortices

Rossby [3] selected long-lived quasi-stationary planetary-scale structures (atmospheric centers of action - ACA) after filtering out pressure field fluctuations related with nonstationary cyclones. To simulate ACA-type modes linear solution of absolute vorticity ($\omega$) equation was used in [3] in the $\beta$-plane approximation. Corresponding consideration was conducted in [4,7-9] on the rotating sphere of radius $R$. Equation of $\omega$ conservation in spherical coordinates ($r, \theta, \varphi$) is in this case the following [10] (for $r \approx R$):

$$\frac{D\omega}{Dt} = \frac{\partial \omega}{\partial t} + \frac{V_\theta}{R}\frac{\partial \omega}{\partial \theta} + \frac{V_\varphi}{R \sin\theta}\frac{\partial \omega}{\partial \varphi} = 0. \qquad (1)$$

---

[*] Corresponding author. mokhov@ifaran.ru

Here $\omega = \omega_r + 2\Omega\cos\theta$, $\Omega$ - angular velocity of the sphere rotation (for the Earth $\Omega \approx 7.3 \cdot 10^{-5} \sec^{-1}$), $\theta$ - co-latitude, $\varphi$ - longitude; $V_\theta = R\dot\theta$, $\dot\theta = \dfrac{d\theta}{dt}$, $V_\varphi = R\sin\theta\dot\varphi$, $\omega_r = (\partial(V_\varphi \sin\theta)/\partial\theta - \partial\varphi/\partial\varphi)/(R\sin\theta) = -\Delta\psi$ - radial component of the local vortex field on the sphere, $\Delta$ - Laplace operator; $\psi$ - stream function for which $V_\varphi = -(\partial\psi/\partial\theta)/R$, $V_\theta = (\partial\psi/\partial\varphi)/(R\sin\theta)$ in (1). Equation (1) in the more general case corresponds to the potential vortex conservation, $D(\omega/H)/Dt = 0$, where $H$ - thickness of the fluid layer (see [10]). Equation (1) holds not only for the constant $H$, but also when $H$ is a Lagrange invariant and velocity field (with zero radial velocity component $\dot r = V_r = 0$) is non-divergent, i.e. $divV = \dfrac{1}{R\sin\theta}(\dfrac{\partial V_\theta \sin\theta}{\partial\theta} + \dfrac{\partial V_\varphi}{\partial\varphi}) = -\dfrac{1}{H}\dfrac{DH}{Dt} = 0$. The latter allows to introduce the stream function $\Psi$. Note that relative (local) vorticity $\omega_r$ varies only for the fluid element moving with the change of latitude. Equation (1) is applicable to the flows with any character length scale $L > H$ for $H \ll R$ [6].

Solution of equation (1), obtained in [4], as well as in [3], in linear approximation (particularly for description of ACA type modes in the Northern Hemisphere – Icelandic and Aleutian Low, and also Azores and Hawaiian High) has the following form

$$\psi = \alpha R^2 \cos\theta + \Psi_{\alpha 0}\cos(\beta t + m\varphi)P_n^m(\cos\theta) , \qquad (2)$$

$$\beta = m(\dfrac{2(\alpha+\Omega)}{n(n+1)} - \alpha), \qquad (3)$$

where $n$, $m$ - integers, $m \leq n, n = 1, 2, ..$; $P_n^m$ - adjoint Legendre polynomials. It was assumed in [4] the disturbance amplitude $\Psi_{\alpha 0}$ smallness with respect to the value $\alpha R^2$, characterizing zonal flow intensity $V_{0\varphi} = \alpha R\sin\theta$, relative to which wave disturbance of the vortex field is considered on the base of (1).

In [8,9], it is shown that solution of type (2), (3) preserves its form also for the general case of nonlinear waves when condition $\Psi_{\alpha 0} \ll \alpha R^2$ already is not necessary. Solution (2), (3) in the limit $\alpha \to 0$ describes wave running from east to west with angular speed $c = \dfrac{d\varphi}{dt} = -\dfrac{2\Omega m}{n(n+1)}$. Steady-state conditions of such a wave necessary for description of ACA type modes are found to be possible for definite value $\alpha$ and corresponding zonal flow intensity. The latter is usually related to the meridional temperature gradient and ACA description on the base of (2), (3) is possible only with an additional determination of the value of the free parameter $\alpha$, which is not defined in hydrodynamic model (1). As a result, pure hydrodynamic description of quasistationary structures of ACA type is not possible on the base of solution of (1) in the form (2), (3). Solution of this problem is proposed in Section 3.

In [4], solution of the nonlinear equation (1) is obtained generalizing solution (2), (3) (for the case $\alpha = 0$ in (2), (3)) [8,9]. Meanwhile contrary to (2), (3) the axis of covering all atmosphere global-scale vortex does not coincide as in (2), (3) with the sphere rotation axis. Vortex axis precesses round the sphere axis with rotation from east to west with angular speed $c_1 = -\dfrac{2\Omega}{n(n+1)}$ ($\Omega$ - the Earth rotation frequency). Given in [4] solution takes for $\alpha = 0$ and after substitution of $\cos\theta$ by $\cos u_0 = \cos\theta\cos\theta_0 + \sin\theta\sin\theta_0\cos(\varphi - \varphi_0)$ in (2), (3) the following form:

$$\psi =$$
$$A_1 + B_1[P_n(\cos\theta_0)P_n(\cos\theta) + 2\sum_{m=1}^{n}\frac{(n-m)!}{(n+m)!}P_n^m(\cos\theta_0)P_n^m(\cos\theta)\cos m(\varphi-\varphi_0 + \frac{2\Omega t}{n(n+1)})] =$$ (4)
$$A_1 + B_1 P_n(\cos u_0).$$

Here $\theta_0, \varphi_0$ - spherical coordinates of the initial axis position for the introduced by Barrett planetary vortex [4] in correspondence with observations [11]. According to [11] circulation of the middle troposphere possesses remarkable eccentricity with the center of symmetry of the flow at a considerable distance from the geographical pole. In such a case the intensity of the zonal component of the flow (with respect to the geographic pole) is low, while the meridional component is large. Harmonic analysis shows that most of the energy of the meridional motion is concentrated in the first longitudinal harmonic (wavelength 360 degrees of longitude) [4]. According to [4] the circulation with a large meridional component may actually be a rather symmetric zonal vortex with respect to an eccentric pole.

As noted in [4], such eccentric planetary-scale vortices previously were not considered in contrast to vortices (with smaller scale) considered by Rossby [12,13]. For $n = 1$ Barrett planetary vortex axis is static in the absolute coordinate system in which the sphere rotates with angular speed $\Omega$ ($c = -\Omega$).

For the further consideration it is convenient to represent the solution of the equation (1) obtained in [4] in the following generalized form (see also [14]):

$$\psi = Y\left(\cos u_0\left(\theta,\theta_0, \varphi-\varphi_0 + \frac{2\Omega t}{\nu(\nu+1)}\right)\right),$$ (5)

where $Y$ - eigen-function of Laplace operator, i.e. $\Delta Y = -\frac{\nu(\nu+1)}{R^2}Y$. The case of integer eigenvalues $\nu = n$ corresponds to the solution of (4). For arbitrary (including complex) $\nu$ the solution of similar type was analyzed in [14] in relation with modeling of blockings on the base of more local dipole vortex structures (modons) constructed with the use of functions $Y$. In [14] the case was investigated for modeling of local vortices (modons) with $Y$ in (5) defined via Legendre functions of the first and second kind: $P_\nu^\gamma, Q_\nu^\gamma$, i.e.

$$Y(\theta,\varphi) = G(\varphi)H(\theta), G = e^{\pm i\gamma\varphi}, \gamma = m, m = 0,\pm 1,..., H(\theta) = \begin{cases} P_\nu^\gamma(\cos\theta) \\ Q_\nu^\gamma(\cos\theta) \end{cases}.$$

It is worth to note that for a stationary case with $\Omega = 0$ (i.e. for static sphere) there exists similar to (5), but even more general form (suggested by E.A. Novikov) of the solution of (1) as $\psi = F(\cos u_0)$, where $F$ - arbitrary function of $\cos u_0(\theta,\theta_0,\varphi-\varphi_0)$ [15]. In [15], in particular, on example of linear function $F$, corresponding to rigid-body rotation it was considered a model of the global pollution transfer in the field of statistical ACA type vortex ensemble.

It is possible to get modification of solution (5) in which precession frequency of the Barrett vortex axis may be already independent from eigenvalue $\nu$. Let us represent solution of (1) as linear superposition (in the most general form when $\alpha \neq 0$) of the stream function $\psi_0$ and (5):

$$\psi = \psi_0 + Y + \alpha R^2 \cos\theta$$ (6)

Function $\psi_0$ in (6) corresponds to zero absolute vorticity $\omega = -\Delta\psi_0 + 2\Omega\cos\theta = 0$ and has a form $\psi_0 = -\Omega R^2 \cos\theta$ characterizing rigid-body east to west fluid rotation relative to the sphere surface. In this case the fluid in the absolute coordinate system is static for $\omega = 0$ and $\psi = \psi_0$.

For (6) with any $\nu$ in $Y$ the corresponding global vortices axes rotate with the same angular velocity $\Omega$ from east to west only in the case $\alpha = 0$. In contrast to (5), this velocity does not depend on the value $\nu$ (or $n$). Actually, from (1) and (6) it follows $\dfrac{\partial Y}{\partial t}+c\dfrac{\partial Y}{\partial \varphi}=0$ for $c=\alpha\left(1-\dfrac{2}{\nu(\nu+1)}\right)-\Omega$. When $\Psi_0$ is not introduced as in [9,14], we have $c=\alpha\left(1-\dfrac{2}{\nu(\nu+1)}\right)-\dfrac{2\Omega}{\nu(\nu+1)}$, where $\nu=n$ in [9]. Let us note that similar expression $c=\alpha\left(1-\dfrac{2}{\nu(\nu+1)}\right)$ was obtained in [2] for $\Omega=0$ and $\nu=n$. In [8] and [4] for $\alpha=0$ and $\nu=n$ the expression for angular velocity has a form $c=-\dfrac{2\Omega}{\nu(\nu+1)}$. It means that for (6) in absolute coordinate system Barrett planetary vortex axis [4] is static (for $\alpha=0$ and $\Psi_0\neq 0$) for any value $\nu$ defining the velocity field symmetry type.

Further, in Section 3, the modeling of interaction of ACA-type structures corresponding to Barrett planetary vortices is proposed. New approach for description of point vortices on the rotating sphere allows to do it on the basis of model corresponding to the equation of hydrodynamics on the sphere (1).

## 3. Exact weak solution for the absolute vortex equation

It is possible to search weak solution of the equation (1) in the form of $N$ pairs superposition of point diametrically conjugated vortices on the rotating sphere, hereinafter referred to as antipodal vortices (APV) [16]:

$$\omega=\dfrac{\Gamma_0}{R^2}(\delta(\theta)-\delta(\theta-\pi))+\sum_{i=1}^{N}\dfrac{\Gamma_i}{R^2\sin\theta_i}(\delta(\theta-\theta_i)\delta(\varphi-\varphi_i)-\delta(\theta-\pi+\theta_i)\delta(\varphi-\varphi_i-\pi)) \quad,(7)$$

where $\delta$ - delta-function. Stream function $\psi$ has a form corresponding to the presence of singularities in $2(N+1)$ points of the sphere:

$$\psi=\psi_0+\dfrac{\Gamma_0}{\pi}Q_0(\cos\theta)+\sum_{i=1}^{N}\dfrac{\Gamma_i}{\pi}Q_0(\cos u_i(\theta,\varphi)), \quad (8)$$

where $Q_0(x)=\dfrac{1}{2}\ln\dfrac{1+x}{1-x}$ - Legendre function of the second kind of zero order and zero power, $\cos u_i=\cos\theta\cos\theta_i+\sin\theta\sin\theta_i\cos(\varphi-\varphi_i)$, and $\theta_i,\varphi_i$ - spherical coordinates of the point vortices which may depend on time.

Obviously, the structure of expression (8) exactly corresponds to the particular case of the stream function (6) for $\alpha=0$ and $m=n=0$. It follows from (7) that equation $\int_0^{2\pi}d\varphi\int_0^{\pi}d\theta\sin\theta\,\omega=0$ holds identically for any values $\Gamma_0$ and $\Gamma_i$, $i=1...N$ and corresponds to the known requirement of equality to zero for the total vorticity on the sphere.

Expression for $\psi$ in (8) takes into account that solution of the equation

$$\omega_{r0}=-\dfrac{1}{R^2\sin\theta}\dfrac{\partial}{\partial\theta}\sin\theta\dfrac{\partial\psi_{0V}}{\partial\theta}=\dfrac{\Gamma_0}{R^2}(\delta(\theta)-\delta(\theta-\pi)) \quad (9)$$

is the function $\psi_{0V} = \dfrac{\Gamma_0}{2\pi} \ln \dfrac{1+\cos\theta}{1-\cos\theta}$. The last term in (8) is obtained on the base of solving equation (9) using symmetry considerations with substitution $\cos\theta \to \cos u_i(\theta,\varphi)$ when vortex axis turns in the direction ($\theta_i, \varphi_i$) from θ = 0.

Vortex field (7) and stream function (8) may be used for getting an exact weak (in the sense of generalized functionals) solution of equation (1). As a result, the form is defined for functions $\theta_i(t)$, $\varphi_i(t)$ which are solutions of the following 2N-dimensional Hamiltonian system of ordinary differential equations ($\dot{\theta}_i = \dfrac{d\theta_i}{dt}$ and so on):

$$\dot{\theta}_i = \dfrac{1}{R^2 \sin\theta_i} \dfrac{\partial \psi(\theta_i,\varphi_i)}{\partial \varphi_i} = -\dfrac{1}{\pi R^2} \sum_{k=1,k\neq i}^{N} \dfrac{\Gamma_k \sin\theta_k \sin(\varphi_i - \varphi_k)}{1-\cos^2 u_{ik}},$$

(10)

$$\sin\theta_i \dot{\varphi}_i = -\dfrac{1}{R^2} \dfrac{\partial \psi(\theta_i,\varphi_i)}{\partial \theta_i} = -\Omega \sin\theta_i + \dfrac{\Gamma_0}{\pi R^2 \sin\theta_i} -$$

$$\dfrac{1}{\pi R^2} \sum_{k=1,k\neq i}^{N} \dfrac{\Gamma_k (\cos\theta_i \sin\theta_k \cos(\varphi_i - \varphi_k) - \sin\theta_i \cos\theta_k)}{1-\cos^2 u_{ik}}$$

(see Appendix). Here $\Gamma_k = const$ for any $\Omega$ and $\Gamma_0$, and $\cos u_{ik} = \cos\theta_i \cos\theta_k + \sin\theta_i \sin\theta_k \cos(\varphi_i - \varphi_k)$. System (10) for $\Omega = 0$ and $\Gamma_0 = 0$ with accuracy up to numerical factor ($\pi$) coincides with the corresponding system in [16], where the system is introduced using kinematic considerations (see [2,17]), but not on the base of exact weak solution of hydrodynamics equation (1). In the case under consideration dissipation processes and energy pumping in the system are treated as not significant or balancing each other. And equations (10), following from (1) and (7), (8), must provide conservation of integral invariants of kinetic energy $\overline{E}$, angular momentum $\overline{M}$ and impulse $\overline{P}$, where line above denotes averaging of respective values over the sphere surface. Noted values related to the mass unit in the rotating coordinate system have the following form: $E = \dfrac{1}{2}V^2$, $M = [r \times V]$, $P = V$, where $V = \dfrac{dr}{dt}$, $r$ – radius-vector in Cartesian coordinate system ($x, y, z$), origin of which is in the sphere center. Rotation of the sphere with frequency $\Omega$ is performed round axis $z$ and radial motion is absent ($V_r$=0).

From definition $\overline{E} = \int_0^{2\pi} d\varphi \int_0^{\pi} d\theta \sin\theta E = \dfrac{1}{2} \int_0^{2\pi} d\varphi \int_0^{\pi} d\theta \sin\theta \psi \omega_r$ it follows $\overline{P} = 0$, and also

$$\overline{E} = \dfrac{1}{8\pi} \sum_{i=1}^{N} \sum_{\substack{k=1,\\ i\neq k}}^{N} \dfrac{\Gamma_i \Gamma_k}{R^2} \ln \dfrac{1+\cos u_{ik}}{1-\cos u_{ik}} + \dfrac{\Gamma_0}{2\pi R^2} \sum_{k=1}^{N} \Gamma_k \ln \dfrac{1+\cos\theta_k}{1-\cos\theta_k} +$$

(11)

$$\dfrac{4\pi}{3} \Omega^2 R^2 - \Omega(2\Gamma_0 + \sum_{i=1}^{N} \Gamma_i \cos\theta_i)$$

$$\overline{M}_z = 2\sum_{i=1}^{N} \Gamma_i \cos\theta_i + 4\Gamma_0 - \dfrac{8\pi\Omega R^2}{3}$$

(12)

$$\overline{M}_x = 2\sum_{i=1}^{N} \Gamma_i \cos\varphi_i \sin\theta_i, \overline{M}_y = 2\sum_{i=1}^{N} \Gamma_i \sin\varphi_i \sin\theta_i \qquad (13)$$

values $\theta_i$ and $\varphi_i, i = \overline{1, N}$ in (11)-(13) are functions of time defining from dynamic equations (10) under corresponding initial conditions. For $\Omega = 0$ and $\Gamma_0 = 0$ all four values (11)-(13) are invariants of the system (10). It may be checked by differentiating (11)-(13) over time taking into account (10). The invariants exactly coincide with the invariants of the dynamical system in [16-18][†]. For $\Omega \neq 0$ or $\Gamma_0 \neq 0$ values (13) are already not invariant, but values $\overline{E}$ и $\overline{M}_z$ are still invariants of (10) (it is true for E only if $\Gamma_0 = const$ and $\Omega = const$). Thus, the sphere rotation or accounting of the polar vortices removes degeneration corresponding to the symmetry related with two invariants (13), existing only for $\Omega = 0$ and $\Gamma_0 = 0$. As a result the system (10) has for $\Omega \neq 0$ or $\Gamma_0 \neq 0$ only two independent integral invariants. Moreover, and for $\Omega = \Gamma_0 = 0$ invariants (13) may be not independent from (11), (12). It gives possibility for existence of nontrivial non-stationary vortex modes (see next section) even for the system with two vortex pairs (N=2), in particular.

Also, for $\mathbf{M}$ in spherical coordinate system we have $M_r = 0$, $M_\varphi = RV_\theta = \dfrac{1}{\sin\theta}\dfrac{\partial \psi}{\partial \varphi}$, $M_\theta = -RV_\varphi = \dfrac{\partial \psi}{\partial \theta}$. And impulse vector $\mathbf{P}$ has components $P_r = 0$, $P_\theta = V_\theta$, $P_\varphi = V_\varphi$ (here as before $r = R$, $V_r = 0$). Hence it follows that for any $\Omega$ and $\Gamma_0$ (including zero) there exists only one independent non-zero average value $\overline{M}_\theta$ (because $\overline{P}_\varphi = -\dfrac{\overline{M}_\theta}{R}$, and $\overline{M}_\varphi = 0$, $\overline{P}_\theta = 0$):

$$\overline{M}_\theta = \pi^2 R^2 \Omega - 2\pi\Gamma_0 \pm 4\sum_{i=1}^{N} \Gamma_i \theta_i - 2\pi\sum_{i=1}^{N} \Gamma_i \qquad (14)$$

However, value (14) is not an invariant of the system (10), which has for any $\Omega$ and $\Gamma_0$ an invariant $\overline{M}_{0z} = \sum_{i=1}^{N} \Gamma_i \cos\theta_i$ not coinciding with $\overline{M}_\theta$. It is interesting to consider conditions of invariance of (14), which are related to the necessity of additional accounting of variability in time for $\Omega, \Gamma_0$ [6].

## 4. Dynamics of two vortex pairs (N=2)

Let N=2 in (10) for any constant value of $\Gamma_1$ and $\Gamma_2$. Then invariant $\overline{M}_{0z}$ has a form

---

[†] Stream function $\psi_1 = \dfrac{\Gamma_1}{4\pi}\ln\dfrac{1}{1-\cos\theta}$ used in [16-18] for unitary singular vortex does not satisfy hydrodynamics equations in contrast to $\psi_{0V}$ from (9) (see (A.1), Appendix). For corresponding $N > 1$ vortices in [17,18] their total intensity shall be equal to zero. As a result, these intensities already can't be defined independent of each other in contrast to intensities for APV. For APV correlation between intensities of a vortex and its antipode is initially is installed in the very invariant structure of APV defining direction of rotation axis for Barrett planetary vortex.

$$\Gamma_1 \cos\theta_1 + \Gamma_2 \cos\theta_2 = c_0. \tag{15}$$

Taking into account (15) system (10) is reduced to the following three equations (first two of them are independent from the third one):

$$\frac{d\theta_1}{dt} = -\frac{\Gamma_2}{\pi R^2} \frac{\sin\theta_2 \sin(\varphi_1 - \varphi_2)}{1 - \cos^2 u_{12}}, \tag{16}$$

$$\frac{d(\varphi_1 - \varphi_2)}{dt} = \frac{\Gamma_0}{\pi R^2}(\frac{1}{\sin^2\theta_1} - \frac{1}{\sin^2\theta_2}) +$$
$$\frac{(\Gamma_1 ctg\theta_2 \sin\theta_1 - \Gamma_2 ctg\theta_1 \sin\theta_2)\cos(\varphi_1 - \varphi_2) + \Gamma_2 \cos\theta_2 - \Gamma_1 \cos\theta_1}{\pi R^2 (1 - \cos^2 u_{12})}, \tag{17}$$

$$\frac{d(\varphi_1 + \varphi_2)}{dt} = \frac{\Gamma_0}{\pi R^2}(\frac{1}{\sin^2\theta_1} + \frac{1}{\sin^2\theta_2}) - 2\Omega +$$
$$\frac{\Gamma_2 \cos\theta_2 + \Gamma_1 \cos\theta_1 - (\Gamma_2 ctg\theta_1 \sin\theta_2 + \Gamma_1 ctg\theta_2 \sin\theta_1)\cos(\varphi_1 - \varphi_2)}{\pi R^2 (1 - \cos^2 u_{12})}. \tag{18}$$

Here $\theta_2$ is defined via $\theta_1$ from (15), and $c_0$ - from initial conditions for $\theta_1$ and $\theta_2$. Existence of energy invariant (11) and independence of equations (16), (17) from (18) allows to integrate the system (16), (17) in quadratures (for $\Gamma_0 = const$ and $\Omega = const$):

$$\int du \frac{A(u)}{(1 + A(u))^2 B(u)} = \frac{\Gamma_2 t}{4\pi R^2} + C_2. \tag{19}$$

Here $u = \cos\theta_1$, $C_2$ - constant of integration, $A = c_e (\frac{1+u}{1-u})^{-\gamma_0} (\frac{1 + c_1 - \gamma_1 u}{1 - c_1 + \gamma_1 u})^{-\frac{\gamma_0}{\gamma_1}}$, $\gamma_0 = \frac{\Gamma_0}{\Gamma_2}$,

$\gamma_1 = \frac{\Gamma_1}{\Gamma_2}$, $c_1 = \frac{c_0}{\Gamma_2}$ $B = [1 - c_1^2 + 2\gamma_1 c_1 u - u^2(1 + \gamma_1^2) + \frac{2(A-1)u(c_1 - \gamma_1 u)}{A+1} - (\frac{A-1}{A+1})^2]^{1/2}$,

and constant value $c_e$ is related with invariant (11) and has a form

$$\ln c_e = \left[\ln\frac{1 + \cos u_{12}}{1 - \cos u_{12}} + \frac{\gamma_0}{\gamma_1}\ln((\frac{1 + \cos\theta_1}{1 - \cos\theta_1})^{\gamma_1} \cdot \frac{1 + \cos\theta_2}{1 - \cos\theta_2})\right]$$ (this expression is defined for t

= 0). Obtained general solution (19) depends on intensity of polar vortices and does not depend on $\Omega$.

In particular, for $\gamma_0 = 0$ from (11) it follows that $\cos u_{12} = c_3 = const$, and from (19) it follows

$$\cos\theta_1 = \frac{c_1(\gamma_1 + c_3)}{1 + \gamma_1^2 + 2\gamma_1 c_3} - a\sin(\frac{2\pi t}{T} + c_4), \tag{20}$$

where $c_4$ - integration constant, $a = \frac{\sqrt{(1 - c_3^2)(\gamma_1^2 + 2\gamma_1 c_3 + 1 - c_1^2)}}{1 + 2c_3\gamma_1 + \gamma_1^2}$,

$T = \frac{2\pi^2 R^2 (1 - c_3^2)}{\Gamma_2 \sqrt{1 + \gamma_1^2 + 2\gamma_1 c_3}}$ (under condition $|a| > \left|\frac{c_1(\gamma_1 + c_3)}{1 + \gamma_1^2 + 2\gamma_1 c_3} - \cos\theta_1\right|$). Value of

frequency $2\pi/T$ in (20) two times exceeds the estimate obtained in [16], where it is noted only qualitative behavior of the system for two APV (and only for $\Gamma_0 = 0$) without full solution as in (19) and (20).

It is possible to estimate oscillation period $T$ for atmospheric vortices taking value of their intensity equal to $\Gamma_1 \approx 10^6 \frac{km^2}{day}$ (similar to [18]): $T \sim 1.5$ years when $c_3 \to 0$ and $|\gamma_1| \to 1$ ($c_3 = 0$ with chord distance $d = R\sqrt{2}$ between ACAs). Note that a value close to $\gamma_1 = -1$ may correspond to the case of two vortex structures like Icelandic Low (with $\theta_1 \approx 25^0$) and Azores High (with $\theta_2 \approx 55^0$), and to the vortex chain of subtropical anticyclones placed symmetrically relative to equator as well (for $\theta_1 \approx 55^0$ and $\theta_2 \approx \pi - \theta_1$), if $\Gamma_1 < 0$ (for $\Omega > 0$).

Thus, we obtained a new exact non-stationary solution describing finite-amplitude vortex dynamics of two vortex pairs on the rotating sphere including additional accounting of polar vortices in (19) (in contrast to conclusions in [18] and next consideration in Section 5, where we analyze only small oscillations near equilibria corresponding to stationary vortex configurations). It may be used for modeling of long-period oscillations of ACAs, in particular. It is possible to check that for $\Omega = \Gamma_0 = 0$ there is relation between invariants (11)-(13), particularly for $N = 2$: $\overline{M}_x^2 + \overline{M}_y^2 + \overline{M}_z^2 = 16\Gamma_2^2(1 + \gamma_1^2 + 2\gamma_1 c_3)$. Existence of such dependence of invariants in (11)-(13) gives opportunity for existence of non-trivial periodic dynamic modes considered here for $N = 2$.

## 5. Stationary vortex modes ($N = 2$) and their stability

Let us consider for the system (16)-(18) conditions of existence and stability of stationary modes corresponding to equilibria of the vortex pairs. Under condition $\dot{\theta}_1 = \dot{\theta}_2 = 0$ in (16) equilibrium is possible either for $\varphi_1 = \varphi_2 = const$, or for $\varphi_1 - \varphi_2 = \pi$. It corresponds in general case to two distinct vortex stationary modes on the sphere. For example, in the case $\varphi_1 = \varphi_2$ from the requirement $\frac{d(\varphi_1 - \varphi_2)}{dt} = 0$ and taking into account (17) the following condition is obtained:

$$\frac{\gamma_0(\sin^2 \theta_{20} - \sin^2 \theta_{10})}{\sin \theta_{10} \sin \theta_{20}} = \frac{\gamma_1 \sin \theta_{10} + \sin \theta_{20}}{\sin(\theta_{20} - \theta_{10})}, \qquad (21)$$

where $\theta_{10}$ и $\theta_{20}$ - stationary values of $\theta_1$ and $\theta_2$. Additional condition for the absence of absolute motion in this case according to equality $\frac{d(\varphi_1 + \varphi_2)}{dt} = 0$ in (18) has the form

$$\frac{2\pi R^2 \Omega \sin \theta_{10} \sin \theta_{20}}{\Gamma_2} = \frac{\gamma_0(\sin^2 \theta_{10} + \sin^2 \theta_{20})}{\sin \theta_{10} \sin \theta_{20}} + \frac{\gamma_1 \sin \theta_{10} - \sin \theta_{20}}{\sin(\theta_{20} - \theta_{10})}. \qquad (22)$$

It is possible to show that for $\gamma_0 = 0$ stationary vortex mode (17), (18) existing for $\gamma_1 = -\frac{\sin \theta_{20}}{\sin \theta_{10}} < 0$, $\frac{\Omega R^2}{\Gamma_2} < 0$, is stable with respect to small disturbances.

For $\gamma_0 \neq 0$ considered stationary state is stable with respect to small disturbances only when the following inequality holds:

$$D \equiv A\gamma_1^2 + 2B\gamma_1 + C < 0, \qquad (23)$$

where $\theta_{10} \equiv y, \theta_{20} \equiv z$

$$A = \sin^3 y [\sin(2z - y) + \frac{2\sin^2 y \cos z \sin(z - y)}{\sin^2 z - \sin^2 y}], B = \sin y \sin z (\sin^2 z + \sin^2 y)$$

$$C = \sin^3 z [\sin(2y - z) + \frac{2\sin^2 z \cos y \sin(z - y)}{\sin^2 z - \sin^2 y}].$$

Inequality (23) corresponds to the condition of realization for oscillatory mode of small disturbances described by the following system of equations:

$$\frac{d\theta_1^*}{d\tau} = -\Delta\varphi \sin z / \sin^2(z - y), \quad \frac{d\Delta\varphi}{d\tau} = -\frac{\theta_1^* D}{\sin^2(z - y)\sin^2 y \sin^3 z},$$

where $\theta_1^*(\tau)$ - disturbance of stationary state $y$, and $\Delta\varphi(\tau)$ - deviation from 0 for the longitude difference $\varphi_1 - \varphi_2$, $\tau = t\Gamma_2 / \pi R^2$.

In the case of Icelandic Low and Azores High in the Northern Hemisphere over Atlantic Ocean the stability condition (23) for $R_0 = 0$ is reduced for mean values $z$ and $y$ ($z \sim 55°$, $y \sim 25°$) (according to data presented in [19] for the period 1949-2002) to the following restrictions for the ratio of intensities of the vortices (for $\gamma_1 = -|\gamma_1| < 0$): $1.35 < |\gamma_1| < 5.24$. According to (21) for the used values of $y$ and $z$ the value of $\gamma_0$ must have the form $\gamma_0 = 0.3(1.94 - |\gamma_1|)$.

Expression for the period of small oscillations of vortices around equilibrium established according to (21), (22) with $\gamma_1 \neq -1$, $\sin z > \sin y$ and condition (23) has a form

$$T = \frac{2\pi \sin^2(z - y) \sin z \sin y}{\sqrt{-D}}. \tag{24}$$

According to (24) the value $T$ may take arbitrary large values and describe respective long-period small oscillations near equilibrium if $|D| \to 0$. In particular, it may take place for Azores High and Icelandic Low for $|\gamma_1| \to 1.35$ or $|\gamma_1| \to 5.24$.

The noted important role of polar vortex pair in the case $\gamma_0 \neq 0$ is not however so principally significant as in the case $\gamma_1 = -1$ with $\sin z \neq \sin y$ (i.e. for $z \neq \pi - y$) when stationary mode (21), (22) may be realized only in presence of finite valued parameter $\gamma_0$, characterizing relative intensity of the polar vortex pair. For the constant in time value $\Gamma_0$ such stationary vortex is always unstable relative to small disturbances.

## 6. Estimates of variability (stability) of vortex modes from observations

In the frame of developed in the present paper approach of vortex dynamics analysis on the uniform rotating sphere there was not taken into account the non-uniformity of the underlying surface, non-adiabatic processes and other factors characteristic for real system of atmospheric vortices. Nevertheless, it is possible to estimate significance of contribution of dynamical component of vortex interactions into evolution of large-scale cyclonic and anti-cyclonic ACAs playing important role in the Earth climate system [19-21]. A trial to describe ACA dynamics with the use of point vortices was undertaken, in particular, in [18], but without estimation for significance of contribution of different factors on the base of observational data.

One of the factors not taken into account in the proposed above theoretical analysis and which should influence on relative dynamics and stability of ACA vortex system is the mean temperature difference between oceans and continents. It is possible to analyze mutual positions of ACA in the Northern Hemisphere taking into account variations of surface temperature over land and ocean, in particular for winter from CRU data (http://www.uea.ac.uk/cru/data) for 1949-2002.

There was conducted analysis of variability of the distance between ACAs by longitude $\Delta\lambda$ for the pairs of Atlantic (Icelandic Low and Azores High) and Pacific (Aleutian Low and Hawaiian High) ACAs as anomality characteristics of the position and instability of the vortex pairs. As a criterion of anomality (instability) of mutual ACA positioning in the particular winter, we take exceeding by the value Δλ of the corresponding standard deviation with respect to the mean value over all period under consideration (1949-2002). Also we estimated degree of anomaly (instability) of the temperature difference between land and ocean *ΔT* in the Northern Hemisphere with respect to the mean mode according to observations for the same period (see Appendix, Table 1). We estimated probability of correspondence of characteristics of normality (stability) and anomality (instability) for *Δλ* and *ΔT* (Table 2). For the winter seasons (according to data for 54 winters) for North Atlantic ACA pair the probability of such correspondence is estimated by 67%, and for Pacific ACA pair by 70%.

On the base of the proposed theoretical modeling approach, it was conducted corresponding stability analysis of ACA vortex pairs for the period 1949-2002 with the use of estimates of the theoretical parameter $\gamma_1$ (ratio of circulations of cyclonic and anticyclonic vortices in the respective ACA pair) and latitude-longitudinal ACA coordinates from data analysis for pressure fields (see [19]). According to theoretical estimates stability or instability for the vortex pair Icelandic Low – Azores High corresponds with probability 78% to empirical estimates of normality or anomality of its longitudinal position (Δλ), and for the vortex pair Aleutian Low – Hawaiian High with somewhat less probability (54%) according to theoretical model.

Obtained estimates show comparability of dynamical and thermal factors in formation of stability modes or instability of the ACA mutual positioning on the sphere. Horizontal axis on Fig. 1-3 corresponds to the co-latitude *θ* for the anti-cyclonic ACA vortex center, and vertical - for the cyclonic ACA. Crosses on figures characterize mean values of *θ* for respective vortex of the ACA pair and their standard deviations.

## 7. Modeling of blockings and comparison with the results of approximate theoretical approaches (on $\beta$-plane and on the rotating sphere)

In relation to the problem of modeling of atmospheric blockings (see [22,23]) different approaches were proposed for solution of absolute vorticity equation (1) on the rotating sphere [14,24,25] and with an approximation of $\beta$-plane [26, 27].

### 7.1. Comparison with $\beta$-plane approximation in [26,27]

In [26,27], there were obtained estimates for dependence of the motion velocity $V$ of the point-vortex pair on their mutual distance $d$ and on intensity $\Gamma$ of these vortices united in a single dipole vortex structure of modon type (see [14]). To compare with results [26,27], let us consider dynamic stationary solution of the system (16)–(18), which is corresponding to $\varphi_1(t) = \varphi_2(t) = \varphi(t)$, $\theta_1 = \theta_{10} = const$, $\theta_2 = \theta_{20} = \pi - \theta_{10} = const$, and $V = R\sin\theta_{10}\dot\varphi = const$ has a form

$$V = \frac{\Gamma_0}{\pi R\sqrt{1-d^2/4R^2}} - \Omega R\sqrt{1-\frac{d^2}{4R^2}} + \frac{\Gamma_1}{\pi d\sqrt{1-d^2/4R^2}}, \qquad (25)$$

where $d = 2R\cos\theta_{10}$ - chord distance between point vortices having the same (by absolute value) intensities ($\Gamma_2 = -\Gamma_1$ in (16)–(18)).

For $\Omega = \Gamma_0 \to 0$ value $V$ in (25) coincides with obtained in [26,27] dependence on $d$ in the limit $d/2R \ll 1$. At the same time, for finite values of $d/2R$ (even when $\Omega = \Gamma_0 = 0$) there is significant qualitative difference of the non-monotonic dependence of the value $\Gamma_1$ (for fixed velocity $V$) from noted in [26] monotonic dependence on $d$. According to (25) (for $\Gamma_0 = \Omega = 0$) value $\Gamma_1$ reaches its maximum value $\Gamma_1 = \Gamma_{1max} = \pi RV$ when $d = d_{max} = R\sqrt{2}$. Such stationary dipole vortex object moves with the constant velocity from east to west in the case when $\Gamma_1 < 0$, i.e. when anti-cyclonic vortex is shifted to the pole relative to the cyclonic one (that is characteristic for blockings, in particular of the splitting type [22]). From (25) it follows that under the fixed intensity $\Gamma_1$ there can't exist (for $\Gamma_0 = \Omega = 0$) a dynamic stationary mode with $V < V_{min} = \Gamma_1 / \pi R$. In the $\beta$-plane approximation considered in [26,27], such a conclusion can't be obtained though it qualitatively agrees with conclusion [27] about existence of maximal allowed distance between vortices in a pair.

### 7.2. Comparison with approximation taking into account sphere rotation in [24]

For comparison of obtained conclusions on stability for $N = 2$ with conclusions [24], let us introduce a parameter $G = \dfrac{\bar{M}_{0z}}{\Gamma_0} = \dfrac{\Gamma_1}{\Gamma_0} \cos\theta_{10} + \dfrac{\Gamma_2}{\Gamma_0} \cos\theta_{20} = const$ (as $\bar{M}_{0z} = const$). In particular, for $\theta_{20} = \pi - \theta_{10}$ and $\gamma_1 = \dfrac{\Gamma_1}{\Gamma_2} = -1$ it follows that $\dfrac{\Gamma_0}{\Gamma_1} = \dfrac{2\cos\theta_{10}}{G}$. In the case $\theta_{10} < \pi/2$ from the condition of exponential instability $1 + 4\dfrac{\Gamma_0}{\Gamma_1}\cos\theta_{10} < 0$ we get that instability takes place only for $G < 0$ and $|G| < 8\cos^2\theta_{10} \approx 0.08$ (for $\theta_{10} = 84.3°$ [24]). In [24] similar estimate for the condition of oscillatory instability is obtained: $|G| < 0.1$. In the analyzed case the motion to the west with $\omega < 0$ in contrast to [24] (where it is possible only for $G < 0$) may take place with $\Gamma_1 > 0$ for $G < 0$ and $|G| < 4\cos^2\theta_{10}$, while with $\Gamma_1 < 0$ - for all $G > 0$ (and for $G < 0$ if $|G| > 4\cos^2\theta_{10}$). Hence, conclusions about stability of relational stationary state with $\omega < 0$ (interesting for blocking modeling, in particular), are found to be substantially dependent on $\Gamma_0 \neq 0$. As a result the conclusion about existence of a stability region for stationary vortex modes modeling blockings is obtained in contrast to [24]. This opens the possibility (denied in [24]) of using model of point vortices on the rotating sphere for study of blockings. It is interesting to compare obtained theoretical conclusions with results of analysis of blockings from observations taking into account polar vortices.

### 7.3. Comparison with an exact stationary solution of equation (1) in [25]

In [25], it is considered dynamically stationary solution of (1) corresponding to zonal flow with constant angular velocity. And solving of equation (1) is reduced to description of value $Q = \Delta\Psi - p\Psi$ (where $p$ is inversely proportional to the constant angular velocity of zonal flow), represented in [25] as linear superposition of $N$ $\delta$-singular objects. Value $Q$ already does not correspond neither local, nor absolute vorticity on the sphere. Therefore in [25] there is no requirement on strict equality to zero of its integral over the sphere surface in contrast to such a requirement for local and absolute vorticity. Given in [25] condition (10 a,b) only approximately transforms into condition of equality to zero of the $Q$ integral over the sphere. In the local limit for $N = 1$ stream function $\Psi$ (expressed as in [14] via considered in Section 2 Legendre functions of the first and second kind) corresponds (see (1.5) in [25]) to the stream function introduced in [2,17]. The latter (see Appendix) does not meet hydrodynamics equations in contrast to the stream function of APV.

Let us note that obtained in [25] condition for existence of dynamically stationary mode (for which system zonal flow is preserved) for the case of $N = 2$ (see (15) in [25]) exactly agrees with the condition (22). Actually, it so if in (22) to set $\gamma_0 = 0$ (absence of polar vortices) and to assume that $A_1 \sin\theta_1 = \Gamma_1$ and $A_2 \cos\theta_2 = \Gamma_2$, where $A_1$ and $A_2$ - amplitudes in the stream function $\Psi$ (see (5) in [25]). However, due to the absence in the reduced equation (see equation (3) in [25]) of time evolution, the possibility of stability investigation for indicated stationary mode is excluded. Such analysis of stability of the indicated stationary mode is conducted in the present paper (see (23) in Section 5). In contrast to [24] the conclusion is made about possibility of blocking modeling with the use of stable point vortices system under certain conditions for intensity of the polar vortices.

## 8. Conclusions

Considered up to now independently wave and vortex solutions of absolute vorticity equation (1) actually characterize the same hydrodynamic object which can be described by Legendre functions (of the first and second kind), in particular. This object is characterized by dualism wave-vortex when point vortices correspond to the presence in some locations on the sphere of singularity for these functions which are regular for the rest part of the sphere. Zermelo [2] introduced separately regular (with integer positive $\nu = n$) stream functions on the sphere and a stream of a "simple vortex" $\Psi_1$, which corresponds to superposition of an isolated singular (point) vortex and uniformly distributed (over the sphere) vorticity corresponding to rigid-body fluid rotation with constant angular velocity. From the other side, in [2] it is noted that for every value $n$ of the regular stream function it is possible to consider its superposition with some rigid-body rotation with angular velocity depending on $n$. In [2] it was not used equation (1) with taking into account rotation of the sphere with frequency $\Omega$. For $\Omega = 0$ expressions for $c$ obtained for (2) and (6) in Section 2 coincide nevertheless with the one obtained in [2].

In the present paper we show that stream function $\Psi_1$ of a "simple vortex" [2], used later in [17], does not correspond to the solution of hydrodynamics equations in contrast to the APV stream function. And it is found out (see (6) in Section 2 and (8) in Section 3), that also for the stream function of the system of $N$ singular APV (described also by Legendre function $Q_0$) it is allowed superposition with rigid-body rotation characterized by function $\Psi_0$. However, it is found out that such fluid rotation is possible only relative to the sphere surface rotating in the absolute coordinate system. But regarding absolute coordinate system, fluid characterized by stream function $\Psi_0$, is static. As a result, there is a possibility for modeling with the use of the system of $N$ APV of the position of axes of any number $N$ of ACAs (in the form equivalent to the Barrett planetary vortices which also have static axes in the absolute coordinate system – see (6) for $\alpha = 0$). And this is realized in the present paper (more details in the case of $N = 2$).

This work was supported by the Russian Foundation for Basic Research, by the Programs of the Russian Academy of Sciences, by the Russian Ministry for Science and Education, and by the Russian President grant 467.2012.5.

# Appendix A

It is possible to show that the very function $\psi_{0V} = \dfrac{\Gamma_0}{2\pi} \ln \dfrac{1+\cos\theta}{1-\cos\theta}$, meeting equation (9), corresponds also to an exact solution of three-dimensional stationary equations of hydrodynamics of ideal incompressible fluid on the sphere. Actually, in spherical coordinate system (see [10]) stationary hydrodynamics equations of ideal incompressible fluid in the axis-symmetric case with $\dfrac{\partial p}{\partial \varphi} = 0$ and $\dfrac{\partial v_\varphi}{\partial \varphi} = 0$ when $v_\theta = v_z = 0$ have the following form:

$$\frac{v_\varphi^2}{r} = \frac{1}{\rho_0}\frac{\partial p}{\partial r}, \quad \frac{v_\varphi^2}{r}ctg\theta = \frac{1}{r\rho_0}\frac{\partial p}{\partial \theta}, \tag{A.1}$$

where $\rho_0$ - fluid density, and $p$ - pressure ($\rho_0 = const$). It is not difficult to check that (since $v_\varphi = -\dfrac{1}{r}\dfrac{\partial \psi_{0V}}{\partial \theta}$) the very function $\psi_{0V}$ meets (A.1). At the same time, stream function $\psi_1 = \dfrac{\Gamma_1}{4\pi}\ln\dfrac{1}{1-\cos\theta}$ used in [2,16,17] already does not meet (A.1).

## Appendix B

It may be shown that the form (7), (8) corresponds to an exact weak solution of equation (1) under condition that $\theta_i(t)$ and $\varphi_i(t)$ in (7), (8) meet 2$N$-dimensional system of ordinary differential equations (10). After substitution of (7) in (1), multiplication of the result of differentiation by arbitrary finite function $\Phi(\theta,\varphi)$ and integration over all the sphere surface, one gets:

$$\frac{1}{2}\int_0^{2\pi} d\varphi \int_0^{\pi} d\theta \sin\theta \Phi(\theta,\varphi)[\frac{\dot{\Gamma}_0}{R^2}(\delta(\theta)-\delta(\theta-\pi))+$$

$$\frac{1}{R^2}\sum_{i=1}^{N}[(\frac{\dot{\Gamma}_i}{\sin\theta_i}-\frac{\Gamma_i \cos\theta_i \dot{\theta}_i}{\sin^2\theta_i})(\delta(\theta-\theta_i)\delta(\varphi-\varphi_i)-\delta(\theta-\pi+\theta_i)\delta(\varphi-\varphi_i-\pi))+$$

$$\frac{\Gamma_i(-\dot{\theta}_i)}{\sin\theta_i}(\delta(\varphi-\varphi_i)\frac{\partial}{\partial\theta}\delta(\theta-\theta_i)+\delta(\varphi-\varphi_i-\pi)\frac{\partial}{\partial\theta}\delta(\theta-\pi+\theta_i))+$$

$$\frac{\Gamma_i(-\dot{\varphi}_i)}{\sin\theta_i}(\delta(\theta-\theta_i)\frac{\partial}{\partial\varphi}\delta(\varphi-\varphi_i)-\delta(\theta-\pi+\theta_i)\frac{\partial}{\partial\varphi}\delta(\varphi-\varphi_i-\pi))+$$

$$\frac{V_\theta \Gamma_i}{R\sin\theta_i}(\delta(\varphi-\varphi_i)\frac{\partial}{\partial\theta}\delta(\theta-\theta_i)-\delta(\varphi-\varphi_i-\pi)\frac{\partial}{\partial\theta}\delta(\theta-\pi+\theta_i))+$$

$$\frac{V_\varphi \Gamma_i}{R\sin^2\theta_i}(\delta(\theta-\theta_i)\frac{\partial}{\partial\varphi}\delta(\varphi-\varphi_i)-\delta(\theta-\pi+\theta_i)\frac{\partial}{\partial\varphi}\delta(\varphi-\varphi_i-\pi))]$$

(A.2)

After conducting in (A.2) of integration by parts (taking into account continuity equation $\frac{\partial(V_\theta \sin\theta)}{\partial\theta}+\frac{\partial V_\varphi}{\partial\varphi}=0$) we obtain (taking into account zeroing of expressions near $\Phi(\theta,\varphi)$ and its same derivatives, and arbitrariness of the function $\Phi(\theta,\varphi)$) given system (10) with $\dot{\Gamma}_i=0$ (i.e. $\Gamma_i=const$, $i=\overline{1,N}$) for any $\Gamma_0(t)$. Arbitrariness in defining of the function $\Gamma_0(t)$ can be eliminated by considering the system (15), complementing system (10) in the case when $\Gamma_0$ and $\Omega$ is possible to consider depending on time.

# Appendix C

Table 1. Comparison of vortex pairs characteristics and their instability (stability) for the theoretical model and anomaly (normality) of positions of ACA pairs for different years in winter from observations [19] for the Northern Hemisphere:

parameter $\gamma_1$ for Icelandic-Azores and Aleutian-Hawaiian ACA pairs (defined using data on pressure anomalies in the ACA centers); mean co-latitude $\theta$ for Icelandic (Ic), Azores (Az), Aleutian (Al) and Hawaiian (Ha) ACA;

characteristics of stability (+), instability (-) or position near the boundary of the stability region (0) of Icelandic-Azores and Aleutian-Hawaiian ACA pairs according to the theoretical model;

characteristics of non-anomaly (+) and anomaly (-) of longitudinal position ($\Delta\lambda$) of Icelandic-Azores and Aleutian-Hawaiian ACA pairs, and also for the land-ocean temperature difference ($\Delta T$) (in brackets, estimates of anomaly degree are given: 0 – for deviations from the mean not exceeding one standard deviation (SD), 1 – not exceeding 1.5 SD, 2 – not exceeding 2 SD, 3 – exceeding 2 SD).

| Year | $\gamma_1$ | | Co-latitude $\theta$ (degrees) | | | | Theoretical estimates | | $\Delta\lambda$ | | $\Delta T$ |
|---|---|---|---|---|---|---|---|---|---|---|---|
| | Ic-Az | Al-Ha | Ic | Az | Al | Ha | Ic-Az | Al-Ha | Ic-Az | Al-Ha | |
| 1949 | -1.55 | -0.97 | 27.5 | 42.5 | 35 | 57.5 | + | - | - (1) | + (0) | + (0) |
| 1950 | -2.31 | -1.36 | 27.5 | 57.5 | 40 | 60 | + | + | + (0) | - (1) | - (1) |
| 1951 | -1.49 | -1.83 | 27.5 | 55 | 35 | 60 | + | + | + (0) | + (0) | - (1) |
| 1952 | -1.79 | -1.38 | 27.5 | 50 | 37.5 | 60 | + | + | + (0) | + (0) | + (0) |
| 1953 | -1.40 | -1.79 | 27.5 | 47.5 | 35 | 60 | + | + | + (0) | - (2) | + (0) |
| 1954 | -1.89 | -1.23 | 27.5 | 55 | 32.5 | 60 | + | 0 | + (0) | - (2) | + (0) |
| 1955 | -2.16 | -1.23 | 22.5 | 62.5 | 37.5 | 57.5 | + | + | - (1) | + (0) | + (0) |
| 1956 | -1.94 | -1.28 | 32.5 | 50 | 40 | 60 | + | + | + (0) | - (2) | - (1) |
| 1957 | -2.25 | -1.42 | 30 | 57.5 | 42.5 | 55 | + | + | + (0) | + (0) | + (0) |
| 1958 | -2.07 | -3.10 | 22.5 | 50 | 35 | 62.5 | + | + | + (0) | - (2) | + (0) |
| 1959 | -2.05 | -1.65 | 32.5 | 60 | 37.5 | 60 | + | + | + (0) | + (0) | + (0) |
| 1960 | -1.9 | -2.00 | 32.5 | 60 | 35 | 60 | + | + | + (0) | + (0) | + (0) |
| 1961 | -1.90 | -1.97 | 27.5 | 55 | 37.5 | 60 | + | + | + (0) | + (0) | + (0) |
| 1962 | -1.46 | -1.27 | 27.5 | 60 | 42.5 | 55 | + | + | + (0) | + (0) | + (0) |
| 1963 | -1.73 | -2.30 | 17.5 | 62.5 | 47.5 | 57.5 | 0 | - | - (3) | + (0) | + (0) |
| 1964 | -1.71 | -1.77 | 35 | 42.5 | 37.5 | 55 | - | + | - (2) | - (1) | + (0) |
| 1965 | -1.66 | -1.54 | 17.5 | 42.5 | 40 | 60 | + | + | - (2) | + (0) | + (0) |
| 1966 | -2.93 | -2.28 | 35 | 60 | 35 | 60 | 0 | + | + (0) | + (0) | + (0) |
| 1967 | -1.82 | -1.89 | 27.5 | 52.5 | 37.5 | 60 | + | + | + (0) | + (0) | - (1) |
| 1968 | -1.52 | -1.94 | 17.5 | 52.5 | 42.5 | 60 | + | + | - (2) | + (0) | + (0) |
| 1969 | -0.93 | -2.74 | 20 | 42.5 | 35 | 62.5 | + | + | + (0) | - (1) | - (3) |
| 1970 | -1.74 | -3.05 | 27.5 | 55 | 40 | 62.5 | + | - | + (0) | + (0) | + (0) |
| 1971 | -1.68 | -0.84 | 32.5 | 60 | 32.5 | 55 | + | - | + (0) | - (3) | + (0) |
| 1972 | -1.98 | -0.97 | 27.5 | 42.5 | 37.5 | 55 | + | - | - (2) | + (0) | - (1) |
| 1973 | -1.94 | -2.01 | 27.5 | 57.5 | 37.5 | 60 | + | + | + (0) | - (1) | + (0) |

| Year | $\gamma_1$ | | Co-latitude $\theta$ (degrees) | | | | Theoretical estimates | | $\Delta\lambda$ | | $\Delta T$ |
|------|-------|-------|------|------|------|------|---|---|-------|-------|-------|
| 1974 | -2.10 | -1.48 | 27.5 | 57.5 | 42.5 | 60   | + | + | + (0) | - (1) | + (0) |
| 1975 | -2.05 | -1.34 | 27.5 | 50   | 40   | 60   | + | + | + (0) | + (0) | + (0) |
| 1976 | -1.84 | -1.39 | 17.5 | 47.5 | 37.5 | 55   | + | + | - (1) | + (0) | + (0) |
| 1977 | -1.47 | -3.02 | 35   | 60   | 37.5 | 60   | + | 0 | + (0) | + (0) | + (0) |
| 1978 | -2.10 | -4.04 | 30   | 62.5 | 37.5 | 62.5 | + | - | + (0) | + (0) | + (0) |
| 1979 | -2.52 | -1.91 | 37.5 | 62.5 | 35   | 57.5 | + | + | + (0) | + (0) | + (0) |
| 1980 | -2.15 | -3.04 | 30   | 50   | 40   | 65   | + | 0 | + (0) | - (1) | + (0) |
| 1981 | -1.10 | -3.30 | 17.5 | 47.5 | 42.5 | 60   | - | - | - (1) | + (0) | - (2) |
| 1982 | -1.96 | -1.10 | 30   | 60   | 40   | 60   | + | 0 | + (0) | + (0) | + (0) |
| 1983 | -1.46 | -4.61 | 17.5 | 50   | 37.5 | 65   | - | - | - (1) | - (2) | - (1) |
| 1984 | -1.62 | -1.60 | 27.5 | 55   | 42.5 | 60   | + | + | + (0) | - (1) | + (0) |
| 1985 | -2.05 | -1.44 | 30   | 42.5 | 40   | 52.5 | 0 | + | - (2) | + (0) | + (0) |
| 1986 | -1.43 | -3.15 | 32.5 | 57.5 | 37.5 | 62.5 | + | 0 | + (0) | + (0) | + (0) |
| 1987 | -1.91 | -3.19 | 30   | 50   | 37.5 | 60   | + | - | + (0) | + (0) | - (1) |
| 1988 | -1.87 | -2.09 | 30   | 57.5 | 37.5 | 60   | + | + | + (0) | + (0) | + (0) |
| 1989 | -1.63 | -1.07 | 27.5 | 50   | 37.5 | 55   | + | - | + (0) | + (0) | - (1) |
| 1990 | -2.66 | -1.35 | 27.5 | 47.5 | 35   | 55   | + | + | - (1) | + (0) | - (1) |
| 1991 | -2.12 | -1.63 | 27.5 | 57.5 | 42.5 | 55   | + | 0 | + (0) | - (1) | + (0) |
| 1992 | -1.38 | -3.68 | 27.5 | 42.5 | 32.5 | 60   | + | 0 | - (1) | - (3) | - (1) |
| 1993 | -1.72 | -3.60 | 17.5 | 42.5 | 37.5 | 62.5 | + | - | - (1) | + (0) | - (1) |
| 1994 | -1.53 | -1.78 | 27.5 | 55   | 40   | 60   | + | + | + (0) | + (0) | + (0) |
| 1995 | -1.80 | -3.02 | 27.5 | 55   | 40   | 62.5 | + | - | + (0) | + (0) | - (2) |
| 1996 | -1.34 | -2.34 | 27.5 | 42.5 | 40   | 60   | + | + | - (2) | + (0) | + (0) |
| 1997 | -1.58 | -1.90 | 27.5 | 45   | 42.5 | 60   | + | + | - (1) | + (0) | - (1) |
| 1998 | -1.88 | -3.40 | 27.5 | 50   | 35   | 65   | + | + | + (0) | - (2) | - (2) |
| 1999 | -1.78 | -1.20 | 27.5 | 55   | 37.5 | 60   | + | + | + (0) | + (0) | - (3) |
| 2000 | -1.55 | -1.52 | 17.5 | 50   | 37.5 | 60   | - | + | - (1) | + (0) | - (2) |
| 2001 | -1.39 | -2.55 | 35   | 60   | 35   | 60   | + | + | + (0) | + (0) | + (0) |
| 2002 | -1.28 | -1.59 | 30   | 50   | 35   | 60   | + | + | + (0) | - (1) | - (3) |

## Appendix D

Table 2. Estimates of probability of synchronous anomality (normality) for longitudinal position of ACA vortex pairs with the land-ocean temperature difference, and also with realization of unstable (stable) vortex pair mode for the theoretical model. States when an ACA vortex pair is on the border of the stability region are treated as unstable. The numbers in brackets correspond to the interpretation of such a boundary positions as stable.

| Vortex pair | Hydrodynamic factors – theory | Effects of the temperature difference land-ocean |
|---|---|---|
| Icelandic-Azores | 0.78 (0.76) | 0.64 |
| Aleutian-Hawaiian | 0.54 (0.54) | 0.58 |

## Appendix E

Figures 1-3 show stability regions (noted by dark-grey color) depending on co-latitude $\theta$ of the center of anticyclonic (horizontal axis) and cyclonic (vertical axis) vortices (in degrees) for different parameters $\gamma_1$, corresponding to the theoretical model in the stability criterion (23). Position of Icelandic-Azores ACA pair (with standard deviations) is also presented at Fig.1-3 for different winters: Fig.1 for 1992 with $\gamma_1$ not corresponding to this year but selected close to -1 to illustrate the case with the small stability region, Fig. 2 for 1988, and Fig. 3 for 1964.

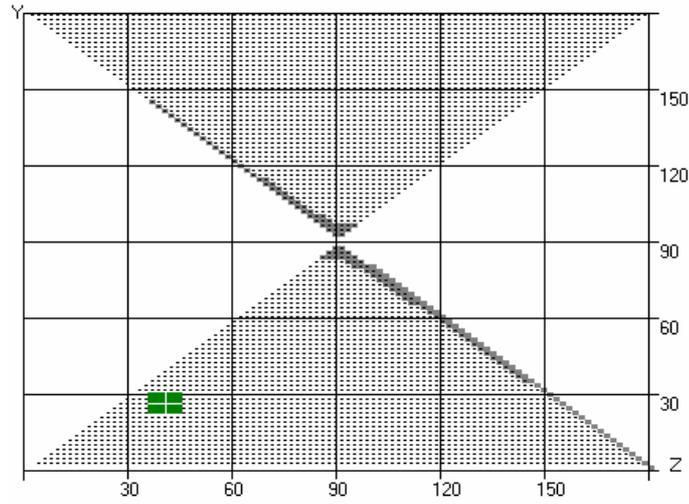

Fig. 1. Stability regions for $\gamma_1 = -1.000067$ with the co-latitudinal position of Icelandic-Azores ACA pair $(27.5^0, 42.5^0)$ for 1992.

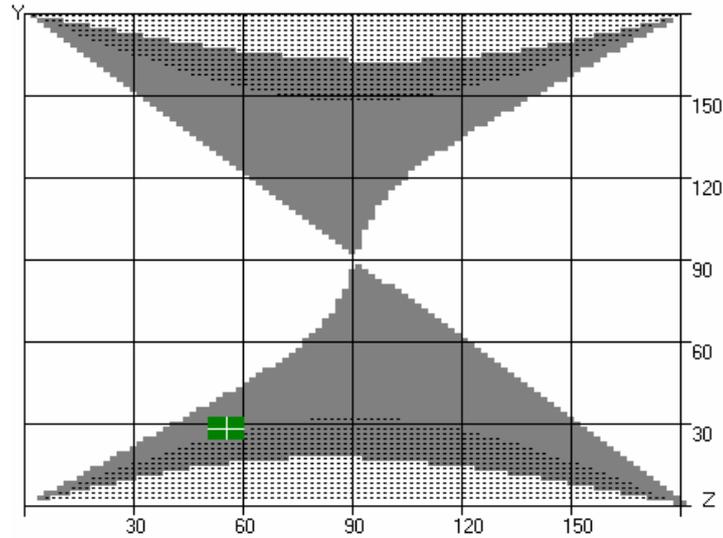

Fig.2. Stability regions for $\gamma_1 = -1.87$ with the co-latitudinal position of Icelandic-Azores ACA pair $(30^0, 57.5^0)$ for 1988.

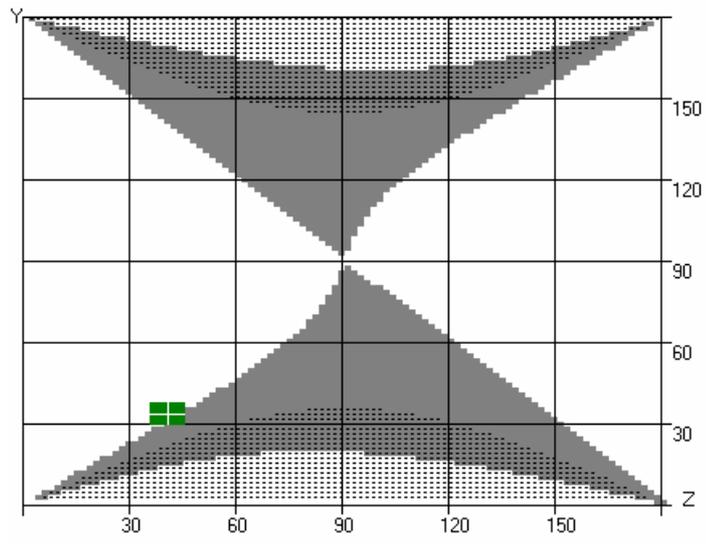

Fig. 3. Stability regions for $\gamma_1 = -1.71$ with the co-latitudinal position of Icelandic-Azores ACA pair ($35^0$, $42.5^0$) for 1964.